\documentclass[journal = nalefd]{achemso}

\usepackage{graphicx} 
\usepackage{color}
\definecolor{Green}{rgb}{0,0.5,0}

\usepackage[english]{babel}
\usepackage[utf8]{inputenc}

\usepackage{amssymb}

\usepackage{ulem}

\usepackage{textcomp}

\newcommand{\mySection}[1]{}

\title{Measurement back-action in stacked graphene quantum dots}
\author{D. Bischoff}
\email{dominikb@phys.ethz.ch}
\author{M. Eich}
\affiliation{Solid State Physics Laboratory,ETH Zurich, 8093 Zurich, Switzerland}
\author{O. Zilberberg}
\affiliation{Institute for Theoretical Physics, ETH Zurich, 8093 Zurich, Switzerland}
\author{C. Rössler}
\author{T. Ihn}
\author{K. Ensslin}
\affiliation{Solid State Physics Laboratory,ETH Zurich, 8093 Zurich, Switzerland}


\date{\today}

\begin{document}



\begin{abstract}
We present an electronic transport experiment in graphene where both classical and quantum mechanical charge detector back-action on a quantum dot are investigated. The device consists of two stacked graphene quantum dots separated by a thin layer of boron nitride. This device is fabricated by van der Waals stacking and is equipped with separate source and drain contacts to both dots. By applying a finite bias to one quantum dot, a current is induced in the other unbiased dot.  We present an explanation of the observed measurement-induced current based on strong capacitive coupling and energy dependent tunneling barriers, breaking the spatial symmetry in the unbiased system. This is a special feature of graphene-based quantum devices. The experimental observation of transport in classically forbidden regimes is understood by considering higher order quantum mechanical back-action mechanisms.

\includegraphics[width=8cm]{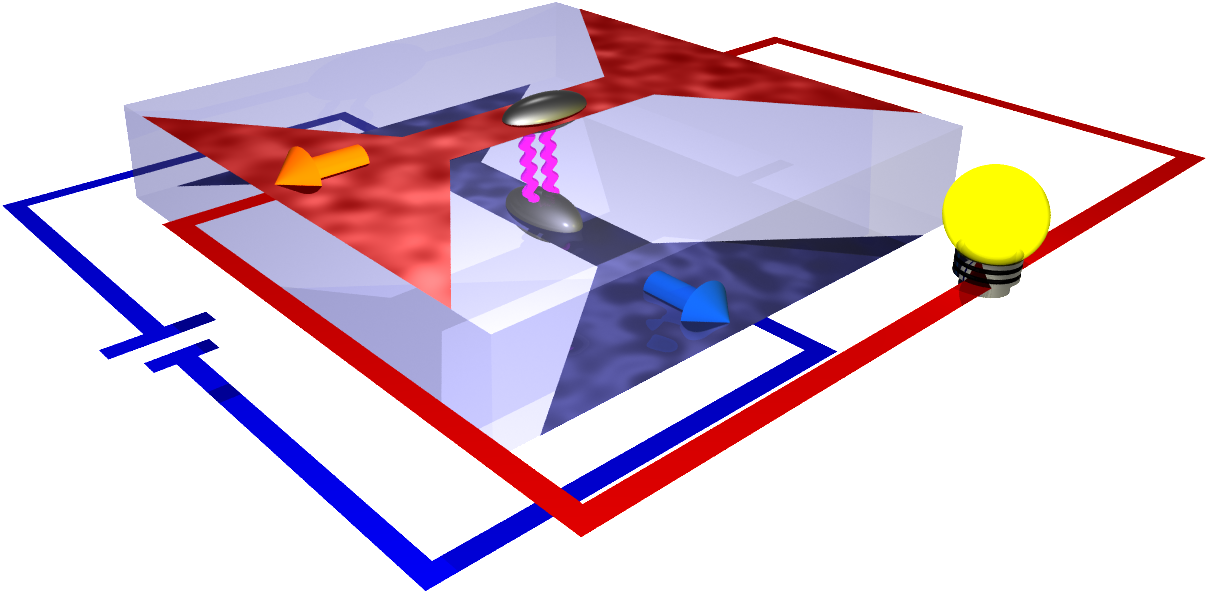}
\end{abstract}



\maketitle



\mySection{Introduction}



A key property of a quantum measurement is that the acquisition of information about a quantum state changes the state itself.~\cite{Neumann1983} This is commonly referred to as ``detector back-action''. An experimental detector is a mesoscopic system that is often well described by a continuous process of gradual information gain alongside gradual back-action.~\cite{gurvitz,Korotkov:2001b,Clerk2010,Field1993,Elzerman2003,DiCarlo2004,Harbusch2010,Gasparinetti2012,Granger2012,clemens2012,clemens2013,zumbuhl2014} Adopting such a description for quantum measurements has enabled the theoretical prediction of the associated detector back-action on many-body virtual transport processes.~\cite{Zilberberg2014}

Experimentally, the electronic properties of a quantum dot are best studied if the coupling to its environment is weak. Coulomb blockade can be observed in electronic transport experiments where the dot is weakly coupled to source and drain contacts via tunneling barriers. By tuning the dot potential with electrostatic gates, the addition-energy levels of the dot can be aligned with the electrochemical potentials of the leads, bringing the dot into a resonant (sequential) tunneling regime.~\cite{Beenakker1991} When the dot level is shifted off resonance by more than temperature (a few $k_B T$), the addition of an electron is forbidden due to Coulomb interaction between the electrons.~\cite{Aleiner2002} In this regime, transport through the dot can only occur via cotunneling processes with a virtual intermediate occupation of the dot.~\cite{Glazman2005}

Whenever a charge detector is coupled to the quantum dot (denoted as system), the coupling involves an energy exchange between the system and the detector. Back-action of the detector on the system can lead to a spectral broadening of the dot energy levels as well as to detector-assisted transport.~\cite{Zilberberg2014} The observation of those higher order effects is difficult as it is only possible if competing processes such as broadening by temperature or tunneling are sufficiently reduced.

Here we present an experiment where the strong capacitive coupling of a quantum dot to its detector allows to investigate detector back-action in detail. This was achieved by separating two graphene nanoribbons by a thin hexagonal boron nitride flake, resulting in a van der Waals heterostructure.~\cite{Ponomarenko2011,Bischoff2015} Quantum dots form spontaneously in both ribbons as a result of disorder.~\cite{Bischoff2015-APR} The nanoribbons can therefore be thought of as quantum dots located on top of each other and interacting via the Coulomb interaction, i.e. as a system and a detector. Such a stacked quantum dot geometry has the advantage of a significantly higher capacitive coupling between the dots compared to geometries where the dots are located next to each other.~\cite{Wiel2003} Quantum dots in a stacked geometry are challenging to fabricate in other material systems such as III-V semiconductors, especially if each dot is required to possess its own pair of contacts.~\cite{Wilhelm2000,Wilhelm2002}. Additionally, the use of graphene and hexagonal boron nitride allows us to reduce the vertical spacing of the dots to the nanometer scale.

In our experiment, we drive a current through the detector quantum dot and find that it induces a current in the unbiased system quantum dot.~\cite{Sanchez2010} We show that this current results from mutual gating of the quantum dots~\cite{Sanchez2010} as well as from additional quantum-mechanical contributions.~\cite{Zilberberg2014} This observation is only possible due to the strongly non-monotonic energy dependence of the quantum dots' tunneling coupling~\cite{Guettinger2011-PRB,Bischoff2013,Bischoff2014} resulting in a spatial breaking of the quantum dot symmetry. Such energy dependent tunneling barriers are readily available in graphene nanodevices.~\cite{Guettinger2011-PRB,Bischoff2013,Bischoff2014}

\begin{figure}[tbp]
	\includegraphics[width=8.46cm]{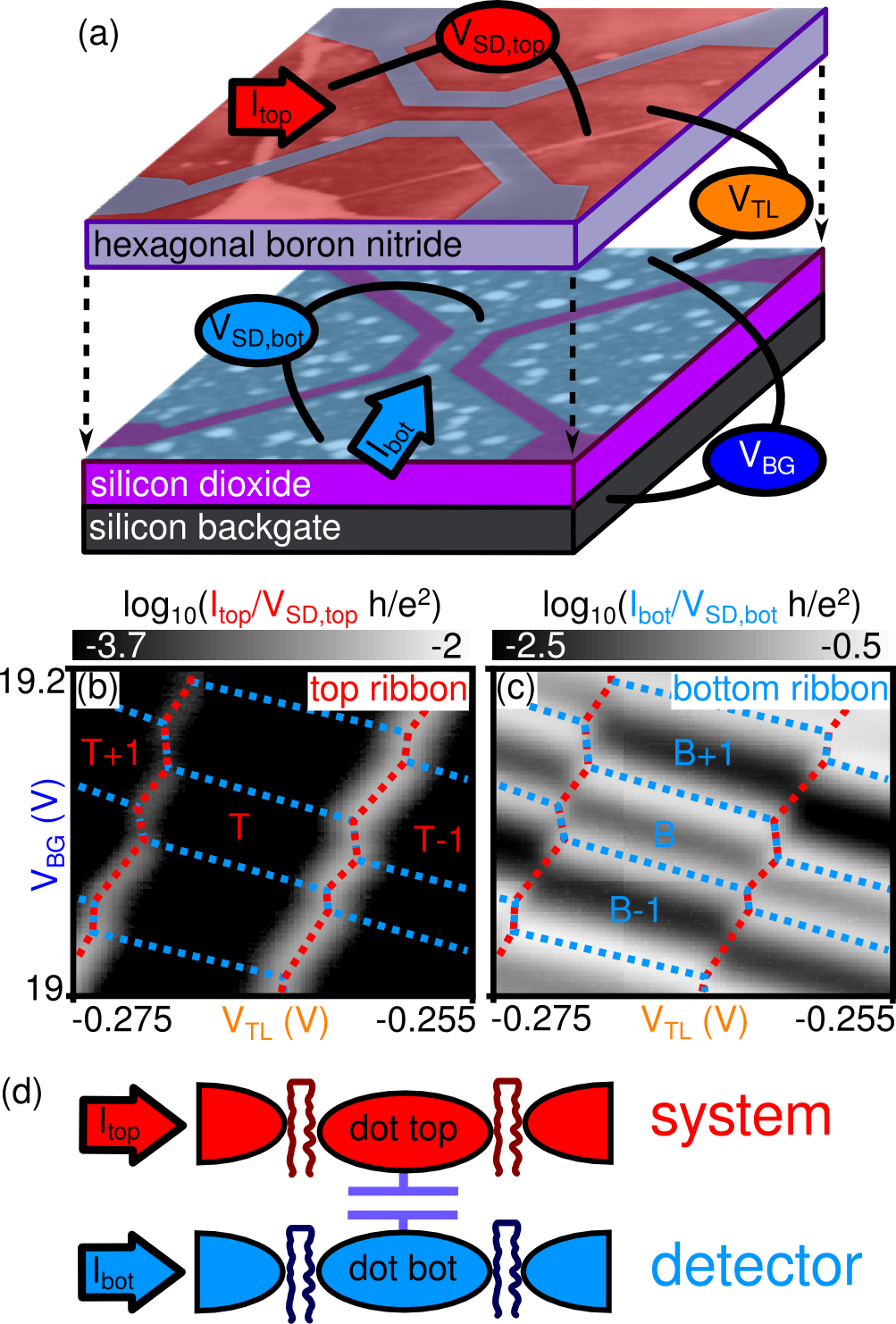}
	\caption{(a) Schematic of the investigated device including false-color scanning force microscopy images: the top nanoribbon (55$\,\mathrm{nm}$ wide, 220$\,\mathrm{nm}$ long) is separated by a hBN flake (12.5$\,\mathrm{nm}$ thick) from the bottom nanoribbon (75$\,\mathrm{nm}$ wide, 200$\,\mathrm{nm}$ long). Applied voltages and measured currents are schematically depicted. (b,c) Conductance of the top/bottom nanoribbon at small bias as a function of $V_\mathrm{BG}$ and $V_\mathrm{TL}$. Coulomb blockade peaks in the top ribbon are marked by red dotted lines and Coulomb blockade peaks in the bottom ribbon by blue dotted lines. These lines form a hexagonal pattern as expected for capacitively coupled quantum dots. The details of the hexagon pattern are however slightly different from typical double dots~\cite{Wiel2003}: the Coulomb blockade peaks in the top ribbon appear with positive slopes. T and B denote the number of charge carriers in the top dot and the bottom dot respectively. (d) Schematic description of the investigated device: a quantum dot in the top ribbon is strongly capacitively coupled to a quantum dot in the bottom ribbon. Tunneling barriers of both dots are distinctly energy dependent, see also Fig.~\ref{Figure2}a.}
	\label{Figure1}
\end{figure}

\mySection{Device and setup}
The investigated device is schematically shown in Fig.~\ref{Figure1}a. It consists of two bilayer graphene nanoribbons separated by a 12.5$\,\mathrm{nm}$ thick hexagonal boron nitride (hBN) flake. The ribbons are aligned perpendicularly to each other and positioned such that they are centered on top of each other. The ribbons were patterned by reactive ion etching (fabrication similar to Refs.~\cite{Dean2010,Bischoff2012}). Each ribbon is independently contacted. The coupling between the two layers is purely capacitive, i.e., no leakage currents from either of the ribbons to any other part of the device can be experimentally detected. All data shown in this paper was recorded at a temperature of 1.3$\,\mathrm{K}$. In the presented data, the bottom graphene layer is kept at a voltage of zero (ground). An additional DC bias is applied between source and drain of the bottom ribbon to drive a current. The top graphene layer is kept at the voltage $V_\mathrm{TL}$ and an DC bias is applied between source and drain of the ribbon in the top layer. Additionally, a global back gate is present ($V_\mathrm{BG}$), that can be used together with $V_\mathrm{TL}$ to tune the electrostatic potentials of the ribbons.

\mySection{Double dot physics}
Electronic transport in both nanoribbons is governed by Coulomb blockade.~\cite{Oezyilmaz2007,Liu2009,Molitor2009-PRB,Todd2009,Bai2010,Han2010-Ribbons} Figures~\ref{Figure1}b,c show Coulomb blockade resonances measured simultaneously in both ribbons at small applied bias voltages and as a function of the gate voltages $V_\mathrm{BG}$ and $V_\mathrm{TL}$, revealing the charge stability diagram of a double quantum dot with capacitive coupling.~\cite{Duncan1999,Wilhelm2000,Chan2002,Wilhelm2002,Holleitner2003,Rogge2004,Huebel2007,Huebel2008}. The coupling capacitance between the two dots is estimated directly from the charge stability diagrams and finite bias measurements to be $C_\mathrm{BT}\approx 10-20\,\mathrm{aF}$~\cite{Wiel2003}. This value is comparable to the value obtained by using a parallel plate capacitor model with the area of overlap between the ribbons and a dielectric constant of $\epsilon_{hBN}=4$. The coupling capacitance per area is already as high as in the most strongly coupled GaAs quantum dots~\cite{Wilhelm2000,Wilhelm2002} and can in principle be significantly enhanced by making the hexagonal boron nitride layer thinner. Further, note that this textbook-like double-dot behavior is only observed for particular gate voltage ranges (see also supplementary materials). Generally, transport in each ribbon is governed by multiple Coulomb blockade resonances with different slopes in the plane of the gate voltages indicating that multiple sites of localized charges contribute to transport. Each ribbon can therefore be understood as a coherently coupled ``multi-dot molecule''.~\cite{Droescher2011,Bischoff2013} This paper focuses on those regimes only where the device behaves as a double dot, as schematically depicted in Fig.~\ref{Figure1}d. Even in those carefully selected regimes, it is likely that additional localized states are present in each ribbon which might contribute to the total current via cotunneling (see also Refs.~\cite{Bischoff2013,Bischoff2014}).

\begin{figure}[tbp]
	\includegraphics[width=8.46cm]{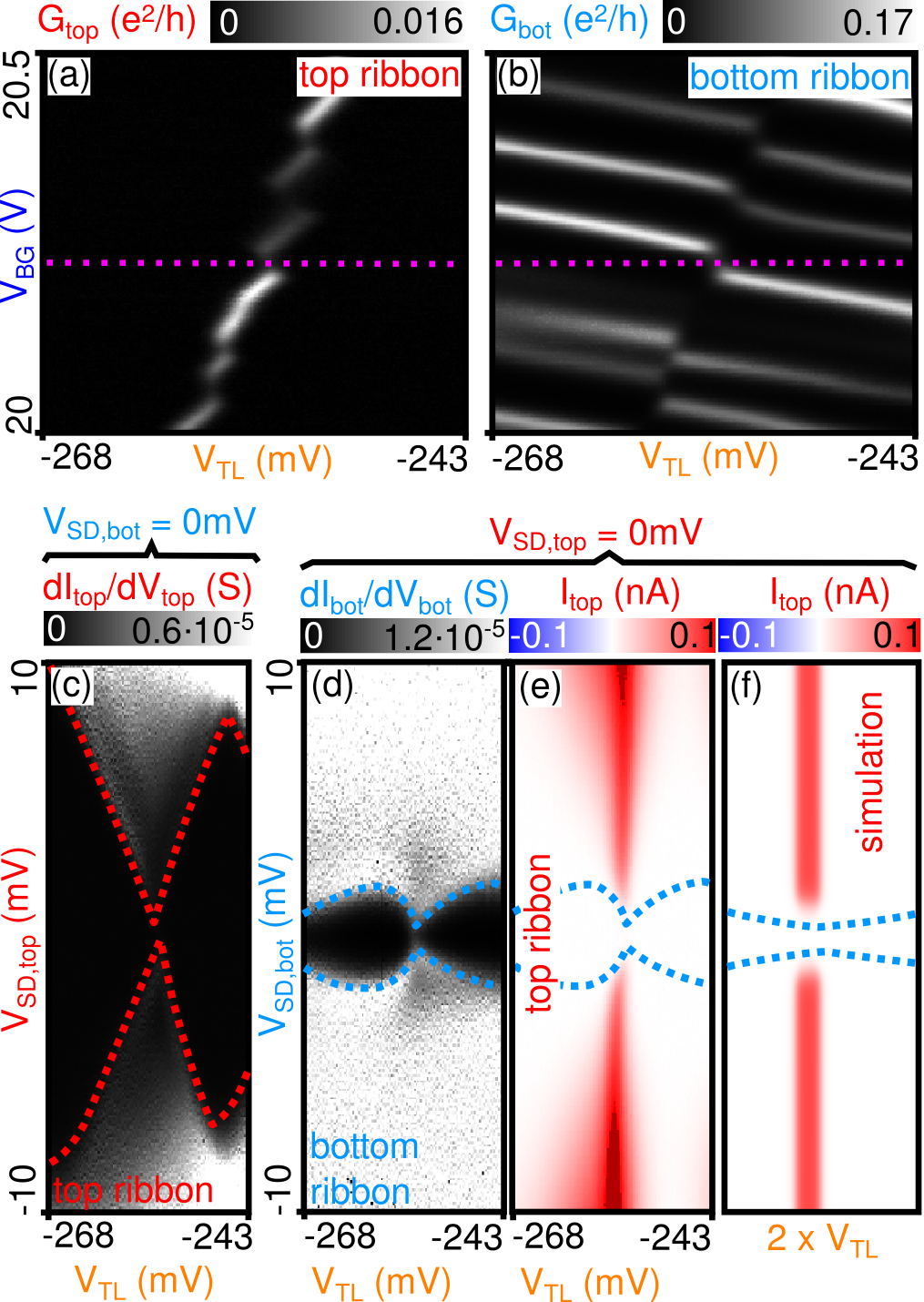}
	\caption{Conductance of (a) the top ribbon and (b) the bottom ribbon at small bias applied to both of them. (c-e) Cuts along the dashed line. (c) Coulomb blockade diamonds in the top ribbon recorded while zero bias was applied to the bottom ribbon. (d) Coulomb blockade diamonds recorded in the bottom ribbon while zero bias was applied to the top ribbon. (e) Current flowing through the top ribbon while the data in (d) was recorded: despite applying zero bias to the top ribbon, a finite current is observed,  depending on the bias of the bottom ribbon. The blue dashed line highlights the outline of the Coulomb blockade diamonds from (d). (f) Corresponding simulated data using the model~\cite{Sanchez2010} depicted in Fig.~\ref{Figure3}. The x-axis is squeezed in order to emulate the more complicated capacitance distribution of the device compared to the model.}
	\label{Figure2}
\end{figure}

\mySection{Energy dependent tunneling barriers}
Following a single Coulomb blockade peak in one of the ribbons by changing the applied gate voltages, the current often decreases and increases in an unpredictable way (see e.g. Fig.~\ref{Figure2}a). This indicates that the tunneling rates change non-monotonically as a function of energy~\cite{Guettinger2011-PRB}, which is well-known for Coulomb blockade in graphene nanoribbons.~\cite{Guettinger2011-PRB,Bischoff2013,Bischoff2014} This behavior is schematically depicted by the energy dependent tunneling barriers in Fig.~\ref{Figure1}d.




\mySection{Nonzero drag currents}
We analyze the device in the framework of ``system'' and ``detector'' because adding an electron to one of the dots can significantly change the current in the other dot. Detector back-action is investigated at the charge degeneracy line of the two dots, as this is the regime where the charge occupancy can simultaneously be changed in both dots. The roles of the two dots as  detector and system can be exchanged, yielding experimentally equivalent results. Figures~\ref{Figure2}a,b show the linear conductance of the system and detector quantum dots in a regime where the system quantum dot is weakly coupled to the leads. Figures~\ref{Figure2}c,d show the corresponding Coulomb blockade diamonds measured for each dot separately along the dashed purple line in Figs. 2a,b crossing the charge degeneracy line, while applying zero bias to the other dot. A gap is observed between the diamond boundaries at positive and negative source-drain voltages due to interdot Coulomb blockade (the gap size is upper-bound by $e^2/C_\mathrm{BT}$). 

In the following, we use the top ribbon as the ``system'' where zero bias is applied and the bottom ribbon as the ``detector'' which is driven out of equilibrium by the applied bias voltage (note that the following discussion is qualitatively also valid if the top ribbon is used as the ``detector'' and the bottom ribbon as the ``system''). For this, we focus on the situation shown in Fig.~\ref{Figure2}d, where zero source-drain voltage is applied to the system dot, whereas the detector dot is driven by a finite source-drain bias voltage. In Fig.~\ref{Figure2}e we see that the non-equilibrium current forced through the detector dot causes a finite current to flow through the system dot in the absence of a voltage drop across it. The observed current flow at zero applied bias is a signature of detector back-action and evidence for broken detailed balance in the system dot.~\cite{Sanchez2010} The current peak observed in the system dot as a function of $V_\mathrm{TL}$ at finite detector bias broadens with increasing detector bias. Depending on the investigated crossing, this current in the system dot can reach up to a few percent of the driving current in the detector dot. We found an induced current in the system dot (independent of the top or the bottom ribbon being used as system) in the majority of the  investigated crossings of Coulomb blockade peaks in the top and the bottom ribbon, as long as current through both ribbons is sufficiently suppressed such that Coulomb blockade diamonds can be measured. The direction in which current flows in the system dot depends on the specific Coulomb blockade resonance at the level crossing of the system dot. Current is found to be positive or negative with equal probability.

Current in the unbiased system dot can only flow if it is driven out of equilibrium by the interaction with the detector, which is itself driven out of equilibrium by the applied bias voltage. This implies energy transfer from the detector to the system dot. However, energy transfer alone is insufficient to generate a directional current through the system dot. An energy-dependent tunnel coupling asymmetry to source and drain is required for giving a preferred spatial direction to the non-equilibrium carriers generated in the system dot.

Two energy-transfer mechanisms can result in back-action: energy transfer via phonons (heat)~\cite{Gasser2009} or via the Coulomb coupling between detector and system dots.~\cite{Shinkai2007,Gustavsson2007} In our system, energy exchange via phonons is suppressed compared to bulk materials due to the stacking of different materials. We further find that the induced current is strongest when both dots are weakly coupled to the leads and capacitive interdot coupling becomes strong on this scale. Contrarily, the situation where the detector ribbon is strongly coupled to the leads resulting in a high current and therefore strong heat generation results in negligible back-action induced currents in the system dot (not shown here). We conclude that energy is therefore mostly transferred via Coulomb coupling.

Before discussing the details of the back-action mechanism, several competing effects are excluded. Conventional Coulomb drag~\cite{Gramila1991,Kim2012,Gorbachev2012,Laroche2011,Laroche2014} relying on momentum transfer between electrons in the two ribbons can be ruled out due to the perpendicular alignment of the ribbons and because inverting the detector bias does not result in a sign change in the system current. Capacitive cross-talk over cables in the measurement setup does not play a role as only DC voltages were applied. DC offsets were carefully corrected and  rectification effects due to noise in the measurement setup were found to be significantly smaller than the observed back-action induced currents.


\begin{figure}[tbp]
	\includegraphics[width=8.46cm]{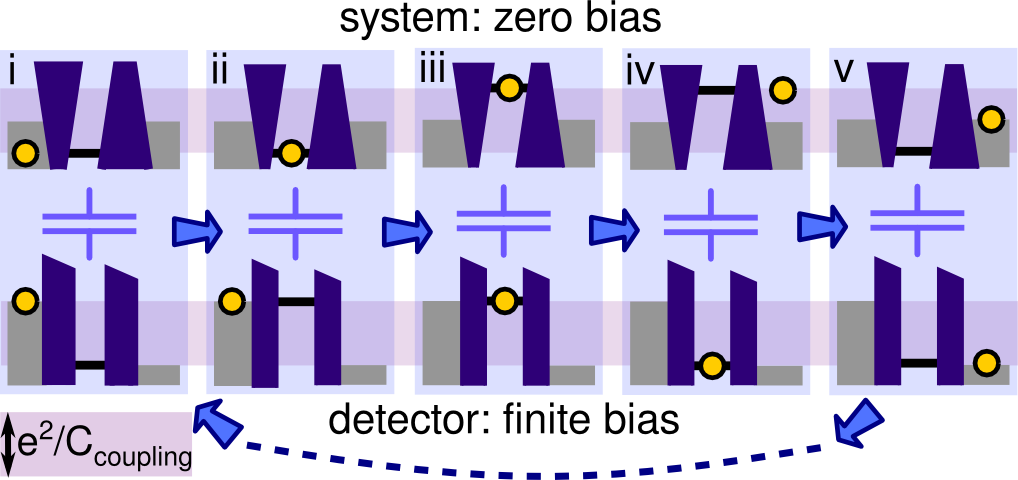}
	\caption{One possible realization of the ratchet-like process.~\cite{Sanchez2010} (i) Both the system dot and the detector dot are empty. (ii) An electron is loaded into the system dot predominantly from the left lead due to the asymmetry of the tunneling barriers. The energy level of the detector dot is shifted up (by $e^2/C_\mathrm{BT}$). (iii) The detector dot is loaded, shifting the energy level of the system dot (by $e^2/C_\mathrm{BT}$). (iv) The electron of the system dot tunnels out, preferentially to the right lead. (v) The detector dot is emptied, resulting in the initial situation. By repeating this cycle, a net current in the system dot arises at zero applied bias. Note that this cycle can for example be aborted if the electron from the detector dot tunnels out before the electron from the system dot leaves the dot. These interrupted cycles together with electrons in the system dot entering/leaving through the slow tunneling barrier limit the amount of current flowing in the system dot.}
	\label{Figure3}
\end{figure}

\mySection{Mesoscopic Coulomb drag}
A system of two strongly capacitively coupled and independently addressable quantum dots with energy-dependent tunneling coupling was theoretically investigated by Sanchez et al.~\cite{Sanchez2010}, who predicted a phenomenon which they called ``mesoscopic Coulomb drag''. The strong Coulomb coupling between the detector and the system together with energy dependent tunneling barriers in the system dot lead to a ratchet-like resonant current generation in the system dot at zero applied bias.~\cite{Sanchez2010} One particular toy model realization of ``mesoscopic Coulomb drag'' is depicted in Fig.~\ref{Figure3}. A few important conclusions can directly be drawn from this model. For a current to flow in the system dot, three conditions need to be fulfilled. Firstly, the energy level of the system dot needs to shift from below the electrochemical potentials of the leads to above (steps ii and iv in Fig.~\ref{Figure3}). The shift in energy of the system dot level due to the addition/removal of an electron to the detector dot is given by $e^2/C_\mathrm{BT}$. Secondly, for current to flow through the detector, its dot level needs to be always located in the bias window spanned by the applied bias voltage, independent of the charge state of the system dot (steps iii and v in Fig.~\ref{Figure3}). The shift in energy of the detector dot due to the addition/removal of an electron in the system dot is given by $e^2/C_\mathrm{BT}$. Thirdly, the tunneling rates in the system dot need to be energy dependent to give the current a preferred direction: note that generally an arbitrary energy dependence of the two tunneling barriers is sufficient (see also Eq.~5 of the paper from Sanchez et al.~\cite{Sanchez2010}).


Figure~\ref{Figure2}f shows the current expected to flow in the system dot as a function of the applied bias in the detector dot based on the model from Ref.~\cite{Sanchez2010}. The maximal width of the vertical red stripe is given by the first requirement (shift of system dot levels below/above lead levels) and the capacitive coupling strength between the used gate and the dot. The second requirement gives the lower bound of $e/C_\mathrm{BT}$ for the observed gap in source-drain direction (see also Figure S1 in the supplementary materials). For this calculation,  the tunneling rates of the detector dot were approximated by $\Gamma_{detector}(V_{SD,detector})=2I_{detector}/e$ because of the experimentally observed multilevel transport. This extension of the model~\cite{Sanchez2010} results only in a minimal increase of the system current with increasing detector bias, as here the tunneling rates of the detector are already faster than the tunneling rates of the system. The tunneling coupling constants of the system dot were assumed to depend only on the charge state of the detector dot~\cite{Sanchez2010} and to be centered around the small bias value extracted~\cite{Houten1992} from Fig.~\ref{Figure2}a. All necessary parameters can be estimated except for the asymmetry of the tunneling barriers of the system dot. For Fig.~\ref{Figure2}f, $\Gamma_\mathrm{sys\to left}^\mathrm{det~empty}/\Gamma_\mathrm{sys\to right}^\mathrm{det~empty}=3=\Gamma_\mathrm{sys\to right}^\mathrm{det~full}/\Gamma_\mathrm{sys\to left}^\mathrm{det~full}$ were assumed, following the description in Ref.~\cite{Sanchez2010}. This particular choice of tunneling rates is compatible with the factor of 3 change in current observed in Fig.~\ref{Figure2}a and with previous experiments.~\cite{Guettinger2011-PRB} An accurate extraction of the tunneling rates is however not possible due to the many unknown parameters.

\begin{figure}[tbp]
	\includegraphics[width=8.46cm]{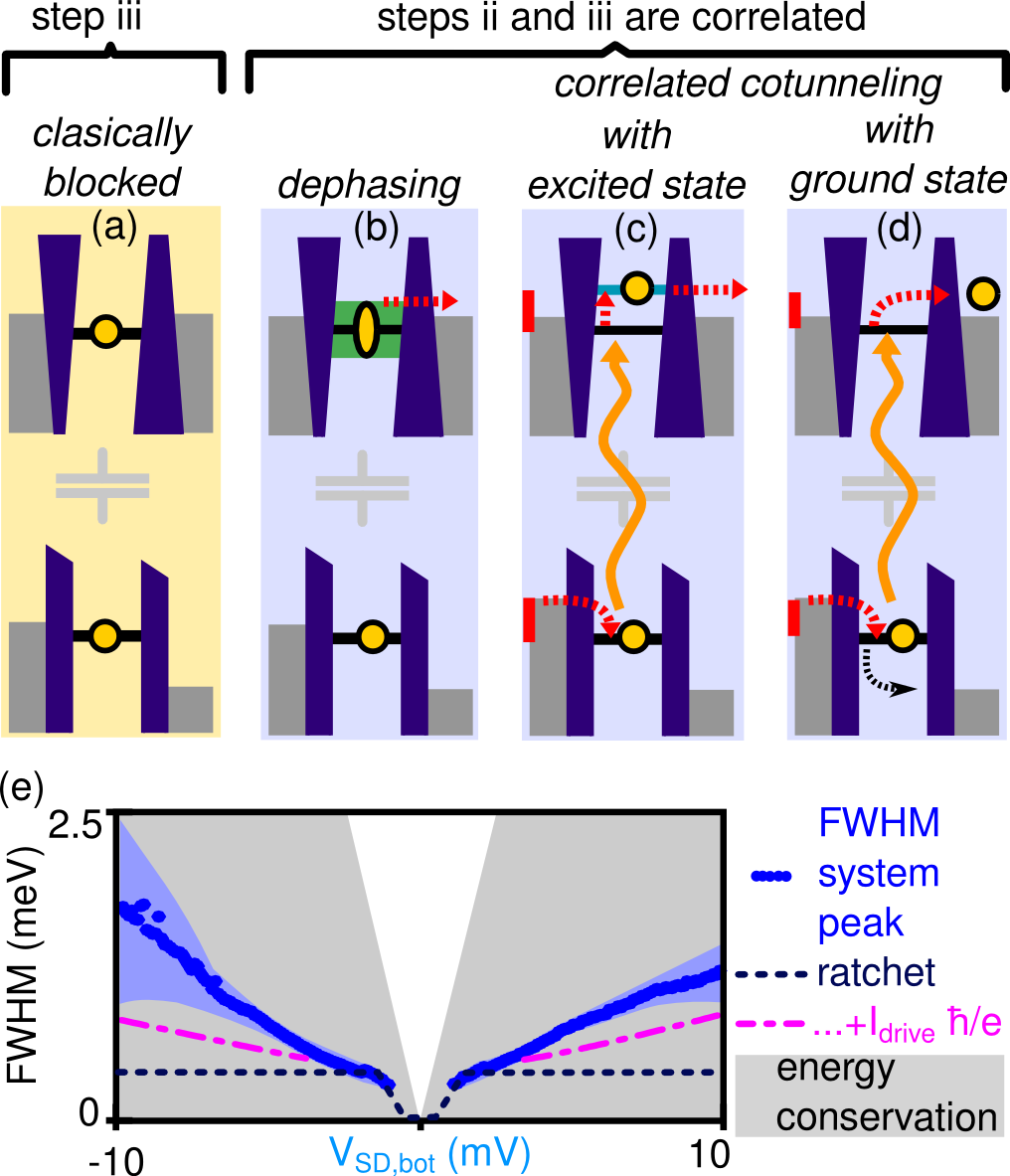}
	\caption{(a) Situation where the gate voltage $V_\mathrm{TL}$ is tuned such that the system dot level is below the electrochemical potentials of its leads even if the detector quantum dot is filled: current is blocked. (b) The current in the detector quantum dot leads to dephasing of the energy levels in the system quantum dot. If the resulting level broadening is sufficiently strong, electrons can tunnel out of the system quantum dot even for the case where the energy level is slightly below the lead levels. (c) Correlated tunneling process where excess energy of the electron tunneling into the detector quantum dot is transferred (orange arrow) to the electron in the system dot and excites it. The excited electron can then tunnel out. (d) Similar to (c) but without the excited state: correlated cotunneling can transfer excess energy from the electron in the detector dot to the electron in the system dot. (e) Blue dots: extracted FWHM of the current peaks in Fig.~\ref{Figure2}e. The blue shaded region illustrates the approximate uncertainty of the extracted value. The gray shaded area depicts the region that is in principle accessible by considering energy exchange. The black dashed line depicts the approximate initial width of the current peak originating from the ratchet mechanism~\cite{Sanchez2010} (including tunneling and temperature broadening). The purple dash-dotted line is offset by the ratchet peak width and depicts a lower bound for the expected broadening of the system dot level due to electrons passing the detector dot.}
	\label{Figure4}
\end{figure}

\mySection{Back-action}
A clear difference is observed between the prediction of the model~\cite{Sanchez2010} from Fig.~\ref{Figure2}f and the experimental data in Fig.~\ref{Figure2}e: at higher detector bias, current starts to flow even outside the region where the resonance condition is fulfilled (first condition not met anymore -- e.g. situation depicted in Fig.~\ref{Figure4}a),  i.e. where transport should classically be suppressed. Considering additional excited states in the system dot that are experimentally present (or any other first order effect) will not change the resonance condition (detailed discussion see also Figure S2 in the supplementary materials). As current starts to flow outside the resonant region (i.e. in the cotunneling regime), we consequently suggest to consider additional higher order quantum mechanical back-action effects as an explanation of the observed data. Furthermore, the current away from resonance increases with detector bias/current, indicating either a broadening of the peak due to the detector back-action or an  additional  energy transfer from the detector to the system. Two classes of mechanisms exist.~\cite{Zilberberg2014} Firstly, the finite lifetime of the system dot level at each specific energy due to the statistical charge fluctuations in the detector dot leads to a spectral broadening of the system dot level (time-energy uncertainty relation). This can be understood as dephasing of the system dot due to the measurement frequency in the detector dot increasing with bias voltage.~\cite{Zilberberg2014,Romito2014,Harbusch2010,Kueng2012,Pulido2014} This lifetime broadening or dephasing occurs on top of the resonant "`mesoscopic Coulomb drag"' mechanism and becomes more important with higher detector current, as depicted in Fig.~\ref{Figure4}b. Secondly, the increased detector bias voltage allows for an energy transfer from the detector to the system additional to the energy (of $e^2/C_\mathrm{BT}$) already transferred by the lowest order process. This results in an increased phase space for electrons in the system dot.~\cite{Zilberberg2014} Many different realizations of this second type of energy transfer exist. For two exemplary processes, the combined and adapted steps ii/iii from Fig.~\ref{Figure3} are depicted in Fig.~\ref{Figure4}c,d (full processes are shown in the supplementary materials, Figure S3). Figure~\ref{Figure4}c depicts a process where an electron with excess energy is tunneling into the detector dot. This excess energy is then transferred to the electron in the system dot, elevating it into an excited state above the electrochemical potential of the leads. Figure~\ref{Figure4}d depicts another option where a correlated cotunneling process transfers excess energy from the electron in the detector dot to the electron in the system dot, allowing it to leave.

To obtain a quantitative comparison between the experimentally observed current and the various theoretical processes, the experimental full width at half maximum (FWHM) of the back-action induced current peak from Fig.~\ref{Figure2}f is shown in Fig.~\ref{Figure4}e as a function of the detector bias voltage. The experimentally found broadening is shown with blue dots and the approximate width of the peak resulting from the lowest order ratchet effect~\cite{Sanchez2010} is marked with a black dashed line. The gray shaded area depicts the criterion of energy conservation assuming that a maximum energy of $eV_\mathrm{SD}$ can be transferred to the system dot.

A lower bound for the level broadening in the system dot due to dephasing is estimated to be $\hbar /\tau_{detector}$, where $\tau_{detector} = |e|/I_{detector}$ is the typical time an electron needs to pass from source to drain in the detector dot.~\cite{Zilberberg2014} This estimate is shown by the purple dash-dotted line in Fig.~\ref{Figure4}b (offset by the lowest order ratchet effect). However, this estimate underestimates the broadening observed in the experiment. Further, this process should conserve the area under the current peak in the system dot~\cite{Zilberberg2014}, which is not observed experimentally. This type of level broadening is therefore insufficient to explain the observed experimental data, indicating that processes similar to Fig.~\ref{Figure4}c,d need to be present as well in order to account for the magnitude of the observed current.

\mySection{Summary and outlook}
In summary, we have shown the fabrication of van der Waals stacked graphene nanostructures. The two ribbons are strongly capacitively coupled to each other without any significant tunneling between them. In selected regimes, the device can be approximated by a parallel double quantum dot with strongly energy dependent tunneling barriers. These settings make it possible to study measurement back-action in detail. We show that at low detector bias, a current starts to flow in the unbiased system dot. This simplest form of back-action is well explained by the model of ``mesoscopic Coulomb drag'', where the mutual gating of the dots together with the energy dependent tunneling coupling leads to a ratchet-like mechanism moving electrons through the unbiased system quantum dot. At elevated detector bias, additional currents start to flow in the system dot beyond resonance in a region where lowest order processes are exponentially suppressed. This is a clear experimental signature of higher order quantum mechanical detector back-action in the regime of many-body correlated described by processes beyond lowest order.

\begin{acknowledgement}
We thank David Sanchez, Gianni Blatter, Florian Libisch, Pauline Simonet, Anastasia Varlet and Hiske Overweg for helpful discussions. Financial support by the National Center of Competence in Research on “Quantum Science and Technology“ (NCCR QSIT) funded by the Swiss National Science Foundation is gratefully acknowledged.  
\end{acknowledgement}



\providecommand{\latin}[1]{#1}
\providecommand*\mcitethebibliography{\thebibliography}
\csname @ifundefined\endcsname{endmcitethebibliography}
  {\let\endmcitethebibliography\endthebibliography}{}


\end{document}